# Design and construction of new central and forward muon counters for CDF II

A. Artikov[1], J. Budagov, I. Chirikov-Zorin, D. Chokheli[2], V. Kolomoets, M. Lyablin, O. Pukhov, A.Sissakian

*JINR, Dubna, Russia*

G. Bellettini, F. Cervelli, M. Incagli, A. Menzione, F. Palmonari, A. Scribano, A. Stefanini

*INFN, Pisa, Italy*

D. Cauz, H. Grassmann, G. Pauletta, L. Santi

*INFN and University of Udine, Italy*

G. Introzzi, A. Penzo

*University of Treste, Italy*

M. Iori

*University of Rome, Italy*

B. Grinyov, V. Lagutin, V. Lebedev, V. Lyubynskiy, V. Senchyshyn

*Institute of scintillating materials, Kharkiv, Ukraine*

V. Kovtun, I. Zaljubovsky

*Kharkiv Karazin National University, Ukraine*

I. Fedorko, S. Tokar

*Comenius University, Bratislava, Slovakia*

V. Giakoumopoulou, N. Giokaris, A. Manousakis-Katsikakis

*University of Athens, Greece*

## Abstract

New scintillation counters have been designed and constructed for the CDF upgrade in order to complete the muon coverage of the central CDF detector, and to extend this coverage to larger pseudorapidity. A novel light collection technique using wavelength shifting fibers, together with high quality polystyrene-based scintillator resulted in compact counters with good and stable light collection efficiency over lengths extending up to 320 *cm*. Their design and construction is described and results of their initial performance are reported.

---

[1] – leave from NPL, Samarkand State University, Uzbekistan.

[2] – leave from HEPI, Tbilisi State University, Georgia.

**Introduction.**

The importance of detecting muons at CDF and of measuring their momentum accurately can hardly be overstressed. Drell-Yan muon-pair production are a means for testing the basics of the EW theory and for searching for new vector bosons, for leptons and quark substructures, as well as for possible extra-dimensions. Muons originate with large branching fractions from the decay of the top quark and of beauty-flavored hadrons. Detailed studies of production and decay properties of the top quark are eagerly awaited and can be best performed in the muon decay channels. An extended beauty physics program, ranging from relatively simple lifetime studies to the study of subtle CP-violation effects can be performed on a muon event sample. A search for the light Higgs boson, that is expected to decay predominantly into b-quark pairs, can be made in the muon b-decay channel. Finally, muons are expected to appear as products of the decay chain in most SUSY processes.

A considerable effort was therefore made in the CDF upgrade to increase the muon acceptance. As a result the refurbished CDF detector that has started a new run of data taking (Run II) in March 2001 features nearly complete muon coverage up to a pseudorapidity $\eta$ of ± 1.5.

Over most of the solid angle the muon detector is a sandwich of drift chambers and plastic scintillation counters, which can be used to trigger on penetrating muon candidates, to identify their bunch crossing and to signal their trajectory. In particular, the fast response of the scintillation counters will be important in associating a muon track with its corresponding bunch crossing.

The complex and compact CDF geometry (see **Fig. 1** and **Ref. 1**) has required the construction and installation of a large set of counter hodoscopes employing a total of about 1200 plastic scintillators with diverse dimensions and light collection assemblies.

Those constructed for the CDF upgrade were designed to overcome the space restrictions and to improve on light collection from long counters. This paper describes the design and performances of these counters, which distinguish themselves by their unconventional and compact light collection system.

The performances of the overall muon detector in terms of muon triggering efficiency and of background rejection will be the subject of a future publication.

**1. Counters for CDF II.**

In CDF, muon momenta are measured by the central tracker in the solenoid magnetic field. However, a key signature is obtained from specialized muon chambers and counters which are deployed behind of the calorimeter so as to exploit the muons' ability to traverse thick absorbers. Counter areas are large in order to keep their number within a manageable limit. Conventional collection systems rely on plexiglas light guides mounted on the counter extremities. These light guides are large and often conflict with other detector parts. Because of this difficulty, the coverage of the Run I muon detector was incomplete. Light collection through the counter extremities also involves a large optical path length in the scintillator and is therefore susceptible to deterioration in the scintillator properties. This susceptibility seriously affected the performance of the Run I counters [2]. The new Run II counters were designed to overcome these problems.

The locations of the CDF II muon detectors are shown in **Fig. 1**. The coverage of the new system in the $\eta$-$\varphi$ space is compared to that of Run I in **Fig. 2**. There are a total of about 1200 muon scintillation counters in the CDF II detector. The longest (up to 320 *cm*) of these counters cover the central ($|\eta|<0.6$) region and are known as the Central Scintillator Upgrade (CSP) counters. They are laid on the outside of the drift chamber of the Central Muon Upgrade (CMP) chamber stacks, behind 61 *cm* of steel shielding located outside the central calorimeter. The new CSP-s cover the top and bottom of the detector (see **Fig. 1**). The older counters cover the north and south walls. Because of deterioration in their transmission properties these older scintillators were refurbished [2] by the addition of a light collection system inspired by the new counter design. The design of the new scintillators, described in this paper, should reduce their susceptibility to this kind of deterioration.

The scintillation counters of the Central Muon eXtension (CSX) are arranged in a conical hodoscope to cover the region $0.6<|\eta|<1.2$. The CSX azimuthal coverage was also completed in the course of the upgrade by the addition of two $90^o$ sectors (the "miniskirts") in the lower parts of both east and west cones [3] and of a $30^o$ sector (the keystone) in the upper part of the west cone. These counters (of the conventional type) had already been designed and constructed before RUN Ib, but their installation had been delayed by several obstructions. The light collection system of a subset of them had to be modified according to the new design in order to overcome these conflicts.

To extend the muon coverage to $|\eta|=1.5$ two overlapping barrel hodoscopes of 160 *cm*-long scintillators (BSU-s) were constructed to cover a barrel of drift chambers (BMU-s) which surround the steel of the former toroid magnets (now decommissioned and used for shielding) for about ¾ of the azimuth (see **Fig. 1, 2**). The bottom of the toroid steel is obstructed by its supports. Finally, a pinwheel counter hodoscope (TSU) is located between the toroids and covers the region from $|\eta|\sim1.2$ to $|\eta|\sim1.5$. Together with the outer BSU ring, these counters form a projective trigger for the BMU-s in the region. The TSU counters were constructed at Michigan State University according to a different design [4].

The design, construction and initial performance of the CSP and BSU counters constructed for the CDF upgrade will be described below. A major improvement in the new counters was achieved by employing ribbons of wavelength shifting (WLS) fibers suitably coupled to the sides of the plastic bars for optimal light collection. Bulky light guides were thus avoided and light could be collected more uniformly and more efficiently from the scintillating material. Light collection from plastic scintillators by means of wavelength shifting fibers was adopted in the past in several sandwich calorimeters [**5, 6, 7**]. This application is different because of the much higher light collection efficiency. In our trigger counters it was vital to reach full efficiency on traversing Minimally Ionizing Particles (MIP). In calorimeters, the amount of light collected from each scintillator layer is not as critical since the detector response is the sum of the signals from a large number of layers.

**2. The R&D stage.**

**2.1. Merits of fiber readout.**

It is customary to collect light from plastic scintillator bars from the short side of the bar with plexiglas light pipes which are either of a "fish tail" type, or an assembly of suitably bent plexiglas strips. However, since the plastic scintillator bars of the CDF II muon detectors are up to 320 *cm* long and their transverse dimensions are relatively small ($\leq 30$ *cm* wide and $\leq 2$ *cm* thick) and light collection through conventional light pipes in counters of this geometry creates several problems:

- The light transmission efficiency in the scintillator is limited by the non-perfect transparency of the bulk material and by the internal reflection efficiency. In our counters both the optical path length and the number of reflections are large. It follows that the plastic surface must be accurately polished to optimize internal reflection and that the bulk transparency must be extremely good. Furthermore, since the counters will be used for many years, both these properties must be time and radiation resistant. Even a small deterioration of the bulk transparency or of the surface quality leads to a large reduction in the amount of light which reaches the photomultiplier (PMT).
- For efficient light collection the area of the PMT photocathode must be as large as that of the scintillator surface which is optically coupled to the light pipe. Since this can be as large as 60 $cm^2$, large and expensive PMT-s would be required for optimal light collection. Moreover, large phototubes are particularly sensitive to stray magnetic fields.
- Large light guides and large PMT-s occupy a lot of space, and overlaps and conflicts with other detectors can be generated.

These problems were addressed and solved to a large degree by the new CDF counter design, which is based on the use of WLS fibers for light collection. This collection method was used for all new counters with mirror differences depending on counter shape. It is schematically shown in **Fig. 3**.

The merits of this method are:

- Light can be collected from counters of any shape with no concern for possible obstructions by light pipes. It is therefore possible for hodoscopes to obtain complete coverage even in complicated geometries and when one is short of space.
- Since light is collected from the long edge of the scintillator bar the number of multiple reflections suffered by the light before exiting the edge of bar is much reduced with respect to conventional light pipes, particularly where the ratio of the transverse to the longitudinal dimensions is small. As a result the tolerance in the quality of bar surface polishing is greatly increased. It also follows that good transparency and light yield of the bulk scintillator become less critical and that scintillator aging and radiation damage can be better tolerated. The overall result is a more reliable and durable detector.
- One is no longer subject to Liouville's theorem and the total cross section of the WLS fibers comprising the light–collecting ribbons, which must be optically connected to the PMT photocathodes is small (~0.5 $cm^2$ in our case). Small PMT-s can therefore be used and the detector can be made more compact with significant cost saving.
- Small PMT-s are less sensitive to magnetic fields. Magnetic shields are often not required or can be much simpler and less expensive.
- The labor costs are greatly reduced with respect to those of light-guide fabrication. When WLS fibers are used one can glue the fiber ribbons to the scintillator and obtain a mechanically solid counter with a relatively simple assembly procedure.
- In some cases, multi-anode PMT can be used to read-out several counters with a corresponding reduction in the cost of the electronic instrumentation.

The design, construction and performances of the new counters are described in the following sections.

**2.2. Manufacturing and testing of prototypes.**

To test the viability of collecting light by means of WLS fiber ribbons as outlined above a number of prototypes of different shapes and sizes were built [8-12] and their light yield was measured.

The study addressed the expected performance of counters with geometries corresponding to the BSU, CSP and WSU[1] counters. The BSU prototype was a rectangular scintillator 180 *cm* long, 17 *cm* wide and 2 *cm* thick. The CSP prototype was 300 *cm* long, 30 *cm* wide and 2 *cm* thick. The trapezoidal WSU counter prototypes were 180 *cm* long and 1.5 *cm* thick, with bases of 40 *cm* and 30 *cm* long.

Counters were wrapped in aluminized paper for diffuse internal light reflectivity and backed by black plastic foil for outside light tightness. An absolute calibration of the PMT (response to single photo-electron (ph.e.)) [13, 14] was obtained by flashing the photocathode with a fast LED. The response of each counter in terms of ph.e./MIP was measured using cosmic muons which were selected by a small counter telescope (4×7 $cm^2$).

The PMT was joined to the fiber ribbon bundle by optical grease. The other end of the fiber ribbon was blackened in a first set of measurement, than grinded flat with sand paper and a small piece of aluminum foil was applied to it with optical glue. This very simply technique provided a reflectivity about 60%. The most relevant results of these tests are shown in **Fig. 4**.

Unlike in the scintillator the WLS fibers are rather transparent to their own emission. Therefore, the effective attenuation length in such counters depends basically on the fiber transparency. The scintillator transparency is of minor importance.

The effective attenuation in the WSU prototype ($\lambda \approx 350$ *cm*) is larger than in BSU prototype ($\lambda \approx 290$ *cm*) due to the configuration of the bars that causes a larger absorption of scintillating light in the WSU counter.

The dopant Y11 features partial overlap of the emission and absorption spectra. The reabsorption effect shifts the emission spectrum to the long wave region, where the absorption is less significant while photons are propagating in the fibers. The partial reabsorption process increases the effective attenuation length in long counters relative to shorter ones. Therefore, the effective attenuation length in the CSP prototype ($\lambda \approx 470$ *cm*) is larger than in BSU ($\lambda \approx 290$ *cm*).

---

[1] The WSU counters were originally intended for the back of the Wall calorimeters and were similar to TSU's, but on the basis of later considerations they were finally not built.

**3. Mass production.**

The scintillator of choice was UPS 923A of Ukrainian fabrication. It was produced at the Institute of scintillating materials in Kharkiv under the supervision of JINR personnel. The bulk material was polystyrene doped with 2% PTP and 0.03% POPOP. The light yield of this scintillator measured in 1992 at the INFN Pisa laboratory [**15**] was found to be superior by ~25% to another commercial scintillator (NE 110) produced at that time. Since then new techniques were implemented at Institute of scintillating materials for polymerizing and machining plastic into long bars [**16**]. The surface machining technique was also changed. After cutting, bar sides were accurately polished until a very smooth surface was reached. The overall light yield from the counter far end (defined as a number of photoelectrons emitted from the cathode when the counter is traversed normally by a MIP) was appreciably improved (**Fig. 5**).

**3.1. Final counters design.**

The final dimensions of the CSP and BSU counters are listed in Table.

Three different types of CSP counters were needed to cover parts of the central detector with different space restrictions. Long counters (CSP) were 2 *cm* thick. The BSU, whose length did not exceed 163.8 *cm,* were made 1.5 *cm* thick. Prototype tests have shown that this was sufficient to ensure efficient counter performances. The typical BSU attenuation curve is presented in **Fig. 6**.

*Table*. Dimensions of the scintillator bars.

| Counters | Length (*cm*) | Wide (*cm*) | Thick (*cm*) |
|---|---|---|---|
| CSP L1 | 240 | 30.5 | 2 |
| CSP L2 | 310 | 30.5 | 2 |
| CSP L3 | 320 | 30.5 | 2 |
| BSU | 163.8 | 16.6 | 1.5 |

A small notch was machined on one corner of the scintillator bars to house the PMT (**Fig. 3**). The area occupied by the notch is only a few tenth of a percent of the total scintillator area, causing only a very small reduction in the geometrical acceptance of the counter. Flat WLS ribbons for light collection were made by gluing 15 or 20 (depending on counter thickness) 0.1 *cm*-diameter fibers, with optical cement (BC 600), directly on the scintillator bar edge. The fiber ribbon protruding into the notch was shaped into a bundle and the bundle was glued inside an adaptor. The ends protruding from the adaptor were cut flat and the surface was grinded and polished as required for good optical contact with the PMT photocathode. In order to maximize light collection the ribbon ends opposite to the PMT were "mirrored". This was accomplished by grinding the flat ribbon end with sand paper and gluing on it a small mirror (see part 3.2).

All faces of the scintillator bars – except the one carrying the WLS fiber ribbon – were accurately polished. Strips of aluminum foils were laid on the side opposite to the ribbon to reflect the light back into the scintillator. A light-reflecting aluminum strip was glued by optical glue to the outer side of the WLS ribbon for increased light capture efficiency.

The large area bar faces were covered with "orange skin" aluminum-backed paper for light reflection.

Blue light is emitted by the wavelength-shifter POPOP in the scintillator. This light reaching the fiber ribbon is absorbed by Y11 (K27 formulation) shifter with subsequent emission in the green region of the spectrum (**Fig. 7, 8**). The green light propagates by total internal reflection in both directions along the fibers. The mirror located on the far edge reflects the light back towards the PMT greatly increasing the light collection efficiency [**8-12**].

Multi-clad S-type WLS Y11 (250 *ppm*) fiber, produced by Kuraray (Japan), was used for most of the WLS fiber ribbons. The fiber core is polystyrene with Y11 shifter, the inner cladding is polymethilmetacrilate (PMMA) and outer cladding is fluorinated PMMA (**Fig. 7**).

Pol.Hi.Tech (Italy) K27 (200 *ppm*) fibers were produced in several different lots. Samples of each lot were quality controlled at the University of Rome. The fibers were excited at different distances from the PMT by light from LED. The LED spectra peaked at 450 *nm* and its intensity was varied changing the LED voltage. During each measurement the LED current was monitored. The 3.3 *m* long fiber was mounted on an aluminum bar and the LED shifted on the bar by a small trolley. For each fiber we took at least 7 measurements between 50 *cm* and 300 *cm* and fitted points by single exponential function. Average attenuation lengths for samples from each lot are shown in **Fig. 9**. The attenuation length of the Kuraray fibers Y11 (250 *ppm*) are seen to be ~20% longer than those of Pol.Hi.Tech fibers K27 (200 *ppm*). Pol.Hi.Tech fibers were used for about 30% of the BSU counters.

It should be noted that the light capture efficiency in multi-clad fibers (~5.35 %, **Fig. 7**) is significantly superior to that of single-clad fibers (~3.14 %). This improvement was essential in bringing the light collection efficiency of WLS fibers into the range which made them useful for the application described here.

The 2.2 *cm* wide, 2.2 *cm* high and 6 *cm* long H5783 photosensor manufactured by Hamamatsu Photonics, which incorporated the 1.5 *cm* wide, 1 *cm* long R5600 PMT, was used as photon detector. The effective diameter of the photocathode of this detector is 0.8 *cm* and its quantum efficiency is of ≈12 % at the average wavelength of the Y11 (or K27) emission spectrum (~500 *nm*).

**3.2. Assembly site.**

About 600 muon scintillation counters were built at JINR for the CDF II CSP and BSU hodoscopes. Although shapes and dimensions were different, the same light collection method as shown in **Fig. 3** was applied to all counters.

During production the counters were subjected to strict quality controls. These included precision in mechanical dimensions, quality of gluing, quality of surface polishing, accurate light shielding and, finally, robust packing for safe shipping. The importance of a very careful gluing of the fragile WLS fibers warrants special mention. Air bubbles must be avoided when gluing because they seriously degrade light collection. Only experienced and adequately trained technical staff was able to perform the work to the required quality.

A custom building-berth was designed and built for gluing fiber ribbons onto scintillator bars. This device allowed to safely gluing ribbons up to 330 *cm* long and up to 31 *cm* wide onto 1.5 and 2 *cm* thick bars. The building-berth had to be flat and smooth to obtain good gluing quality (without bubbles) scintillator-ribbon sandwich. For this purpose one usually makes use of massive granite plates with specially machined surfaces. An original (and considerably less expensive) solution to the problem was obtained by exploiting the simple property of liquids to settle into a flat horizontal surface under the influence of gravity.

The building-berth is shown in **Fig. 10**. It was assembled in a number of steps. First, the berth support frame was firmly anchored on the lab floor. Next, a solid box was created inside it with plastic walls suitably supported by girders on the perimeter. The box was then filled with epoxy diluted with acetone to increase its fluidity. After the epoxy hardened an excellent flat surface was obtained. The maximum deviation from flatness over 4 *m* was < 100 microns.

A 7 *cm* wide Teflon strip was fixed with bolts along the full length of the hardened epoxy surface of the building-berth (**Fig. 11**). A 1.5 *cm* wide groove and a 2 *cm* wide groove were machined along each strip in such a way that departure from perfect flatness did not exceed 50 *microns*. Four such building-berths were built, allowing up to 4 "ribbon/bar" in the CSP case or up to 8 "ribbon/bar" in the BSU case to be glued onto the scintillator bar each day.

Counters were manufactured according to the following procedure. First, a strip of aluminum foil was laid in the groove and held in place with scotch tape. Then, the appropriate number (15 or 20) of fibers, which had already been joined on one end and glued into an adaptor were laid in the groove. The fiber lengths were chosen to be some 5-6 *cm* longer than the ribbon. The adaptor was housed on the counter with the aid of a billet simulating the PMT.

Fibers were carefully stretched out side by side to form a flat ribbon in the groove and a temporary rod was used to keep the ribbon in place at the scintillator ends. Finally, BC-600 optical cement was poured into the groove and spread out evenly until it covered the entire surface of the ribbon. By working on such a flat support the amount of glue needed for gluing a WLS bundle was only ~9 *g* for the BSU and ~21 *g* for CSP counters.

The scintillator bar was then laid on the ribbon and fixed in position at its top edge by means of a clamp. Then nuts were tightened to ~ 6-8 *Newton×m* (**Fig. 11**) for providing the optimal pressure. About 24 hours were needed for the epoxy to harden. Finally the scintillator-ribbon sandwich was carefully removed from the building-berth and positioned on a special table for the further work.

The WLS ribbon was cut at its far end, milled, grinded and polished. A "five-minute" optical epoxy (Devcon, USA) was used to glue the mirror on this ribbon end. The mirror is a ~0.2 *μm* thick *Ag* layer evaporated in vacuum on a plexiglas plate. After about one hour the plexiglas plate could be removed. As a result the *Ag* foil remains glued on the fibers. An aluminum foil was glued over the *Ag* layer as a protection.

The counter surface was then carefully cleaned of the glue spills, fat spots etc. by wiping it first with a soft cloth moistened with warm water. The operation was repeated with a 50% $C_2H_5OH$ (ethyl-alcohol) solution. Finally, the plastic surfaces were wiped dry.

The counter was then moved by hand (cotton gloves were used) to an adjacent table and its large area side was laid on aluminized "orange skin" paper that had been cut to fit the counter shape. The bar edges were covered with aluminum strip so that the counter was fully wrapped in its reflecting coating. Black lightproof ~0.4 *mm* thick plastic sheets were laid on the two large area sides covering them up to 5-6 *mm* from the perimeter. The counter perimeter was then covered by black lightproof 50 *mm* wide electric tape to complete the light isolation of counter. The completed counter was thus ready to be tested with cosmic muons and radioactive sources.

## 4. Counter performance and quality control.

Detectors performance and quality control were studied with cosmic muons. The block diagram of the measurement is shown in **Fig. 12**. The analog signal from the PMT was amplified by a high-speed amplifier Model 777 «Phillip Scientific» (~150 *MHz* band width) and measured by an ADC (Le Croy 2249A). The width of the gate signal was ~80 *ns*.

The PMT signal was suitably amplified to obtain a single photoelectron spectrum of appropriate amplitude for calibration of a spectrometric channel. An attenuator (ATT) was used to increase the dynamic range of measured signals. The ADC output was read and processed by means of a PC. The test setup operated in two regimes with different triggers. Trigger 1 was used to measure the cosmic muon spectra (cosmic muons selected by a telescope of three small 7×7 $cm^2$ scintillation counters S1, S2, S3 in coincidence), and trigger 2 – for calibration of the spectrometric channel.

The counter to be tested was sandwiched between S1 and (S2, S3). Moving the telescope along the counter axis the dependence of the light yield on the distance from the bar edge was measured. Trigger 2 was provided by a pulse generator and was used for the measurement of the light emission diode (LED) spectra. The fast *AlGaAs* LED HLMP8100 ("Hewlett Packard") was driven by a ~10 *ns* long pulse. The photon flux incident on the PMT photocathode was tuned by changing the supply voltage to the LED. The LED spectra were used to determine the spectrometric channel parameters and to monitor their overall time drift. For these purpose measurements of LED spectra we carried out both before and after each cosmic muon run.

The spectrometric channel was calibrated by measuring, in absolute units, the distribution of the number of photoelectrons emitted by the PMT photocathode for a traversing muon.

The light yield in absolute units characterizes the counter performance. In principle, it would be sufficient if it is at least as large as needed to obtain full efficiency, but in practice it is important that it is substantially larger to accommodate degradation with time. The knowledge of light yield in absolute units is very important as it enables not only to find the efficiency of counters and to compare parameters of different detectors, but also to gauge the counters long-term stability [**11, 12, 15**].

The calibration was done by means of a LED using light flashes of low intensity. The basic idea of the calibration method consists in a deconvolution of the LED spectrum with small number of photoelectrons (1-2) using a realistic PMT response function. A typical deconvoluted LED spectrum is shown in **Fig. 13**. It corresponds to an average of µ=1.5 photoelectrons emitted by the PMT photocathode. The solid line shows the PMT response function, with fitted parameters as given in figure. The dashed curves represent the charge distribution for photoelectrons emitted by the first dynode (where can be reached by light because the photocathode is partially transparent). These photoelectrons can be captured by the downstream PMT dynode system. The charge distributions for *n*=1, 2, 3… photoelectrons collected by the PMT photocathode are also shown in the figure. The response function has 8 parameters. Six of them describe the part of spectrum corresponding to the input light signal:

- $Q_0$, $\sigma_0$ – pedestal height and its width;
- $Q_1$, $\sigma_1$ – mean output charge initiated by a single photoelectron created on photocathode and the corresponding standard deviation;
- $K_1$ – secondary emission coefficient of the first dynode;
- $\sigma_2$ – standard deviation output charge when the signal is initiated by a single photoelectron originating from first dynode;
- $\mu$ – mean number of photoelectrons emitted by the photocathode and captured by the PMT dynode system;
- $\mu_1$ – mean number of photoelectrons emitted by the first dynode and captured by the PMT dynode system.

$Q_1$ (in ADC channels per photoelectron) gives the spectrometric channel calibration which is needed to measure the counter light yield (in ph.e./MIP). Details of this method of calibration can be found in **Ref. 17**. Using the calibration constant $Q_1$ the yield dependence on the longitudinal coordinates was measured with cosmic muons. During this measurement the dynamic range of the ADC was adjusted as convenient by varying an attenuator. About two hours long runs were needed for each measurement in order to reach the required statistical precision (~4%).

The calibration measurements were carried out before and after each cosmic muon run and calibration coefficient $\langle Q_1 \rangle$ was found as on average of the two $Q_1$ values [**15**].

A typical cosmic muon spectrum of a CSP counter is shown in **Fig. 14**. The dashed line represents the pedestal centroid $Q_0$. From this spectrum one finds the average muon signal in photoelectrons as:

$$N_{pe} = \frac{\langle Q \rangle - Q_0}{\langle Q_1 \rangle} K_{att},$$

where $\langle Q \rangle$ is an average spectrum amplitude and $K_{att}$ is the attenuation coefficient.

All BSU and CSP counters were tested as described above. **Fig. 15** shows the distribution of photoelectrons over the entire sample for muons traversing close to the far end of counters.

Note that all tests of counters were made without optical grease between bundle ribbon and face of PMT. Using optical grease the light yield can be increased by ≥10%.

**Conclusion.**

A novel technique for collecting light from large area scintillation counters has been developed and successfully applied to the construction of more than 600 new counters, between 160 *cm* and 300 *cm* in length, for the CDF muon upgrade. The technique relies on a wavelength shifter fiber to extract the light from the longer side of the scintillator bar, thereby reducing the path length of the light in the bulk material and consequently the importance of good light transmission in the counter. Performances of long bars are therefore less dependent on the scintillator transmission properties and are less susceptible to its deterioration than with conventional light-guides.

Another important feature of this technique is the reduced cross-section of the fiber bundle, which allows using smaller area phototubes. The elimination of lucite light guides and of large photomultipliers results in a much more compact design for which the ratio of sensitive to total area is close to one. The reduced sensitivity of small photomultiplies to magnetic fields can also be an important feature.

The scintillator used to construct the counters (UPS 923A) was a polystyrene based plastic developed by the Institute of scintillating materials in Kharkiv, under the supervision of JINR and the counters were constructed at JINR using a cost–effective technique which was developed in collaboration with INFN Pisa and finalized at JINR. The results of quality control tests performed at JINR show that the average light–output ranges between 21 *ph.e.*/MIP (for the longer counters) and 28 *ph.e.*/MIP (for the shorter ones) for muons traversing the counters normally at the ends furthest from the photomultipliers. This light collection efficiency is more than adequate for 100% efficiency over the entire counter area. Allowing for a typical deterioration rate of 5–10% per year, full efficiency should be retained for the entire useful lifetime of CDF.

Final assembly and quality control of the counters was performed at FNAL. This procedure and the performance of the counters during data taking will be reported on in a forthcoming article.

**List of figures.**

**References.**

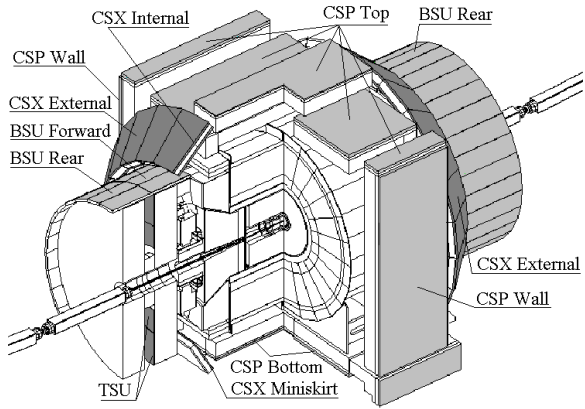

*Figure 1.* The CDF II detector.

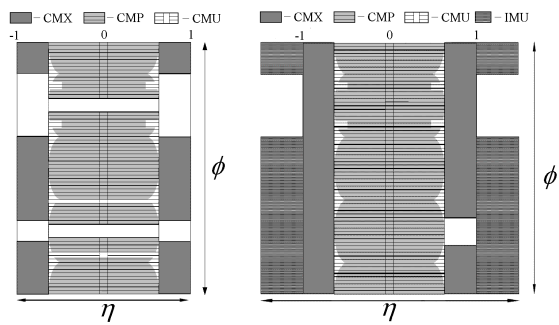

*Figure 2.* Comparison of the muon counter coverage in Run I (left) and in Run II (right), as a function of azimuth $\phi$ and pseudorapidity $\eta$.

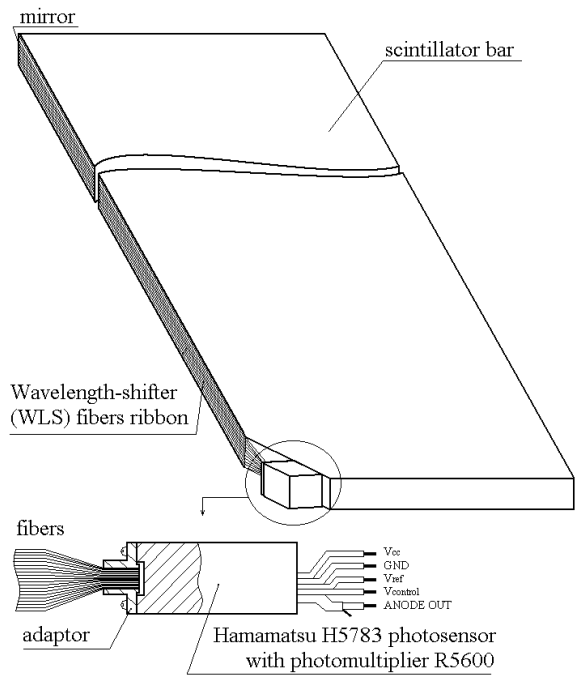

*Figure 3.* Counter construction employing a WLS fiber light guide.

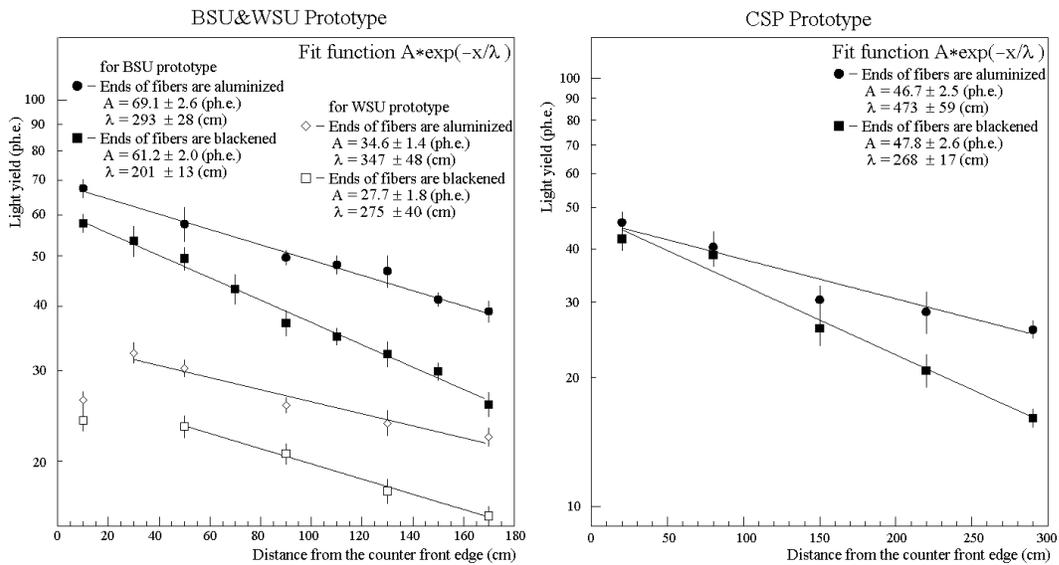

*Figure 4.* Test results for the BSU (180×17×2 $cm^3$), WSU (180×40 (30)×1.5 $cm^3$) and CSP (300×30×2 $cm^3$) counter prototypes.

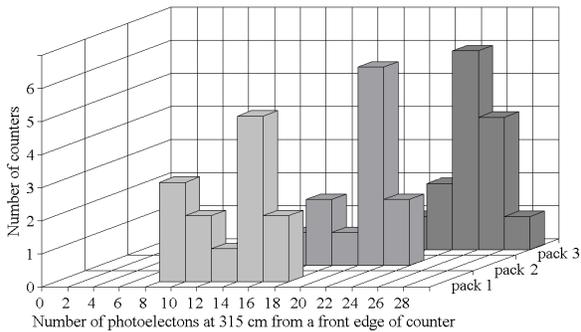

*Figure 5.* Dependence of the light yield from the far end (in ph.e.) of CSP L3 counters from different Kharkiv scintillator lots. The scintillator bars quality improved by about a factor of 2 from lot #1 to lot #3. However, the light yield was in all cases good enough for our purposes.

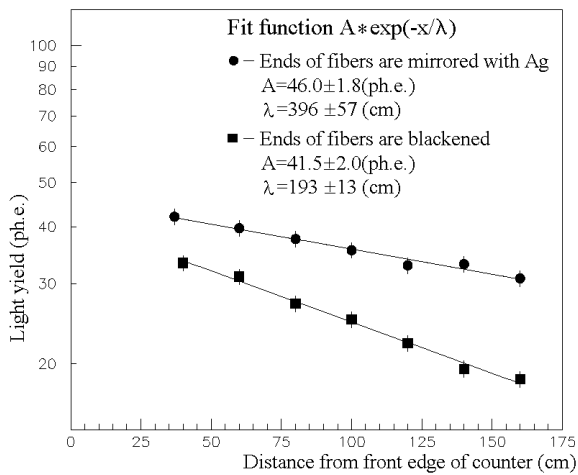

*Figure 6.* Measurements of light yield of a BSU counter (15 WLS fibers, KURARAY).

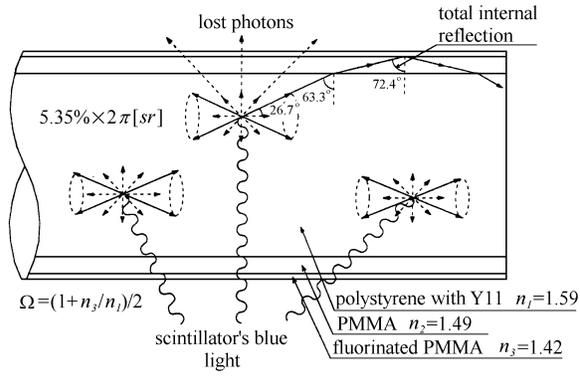

*Figure 7*. Principle of the light collection using WLS fibers.

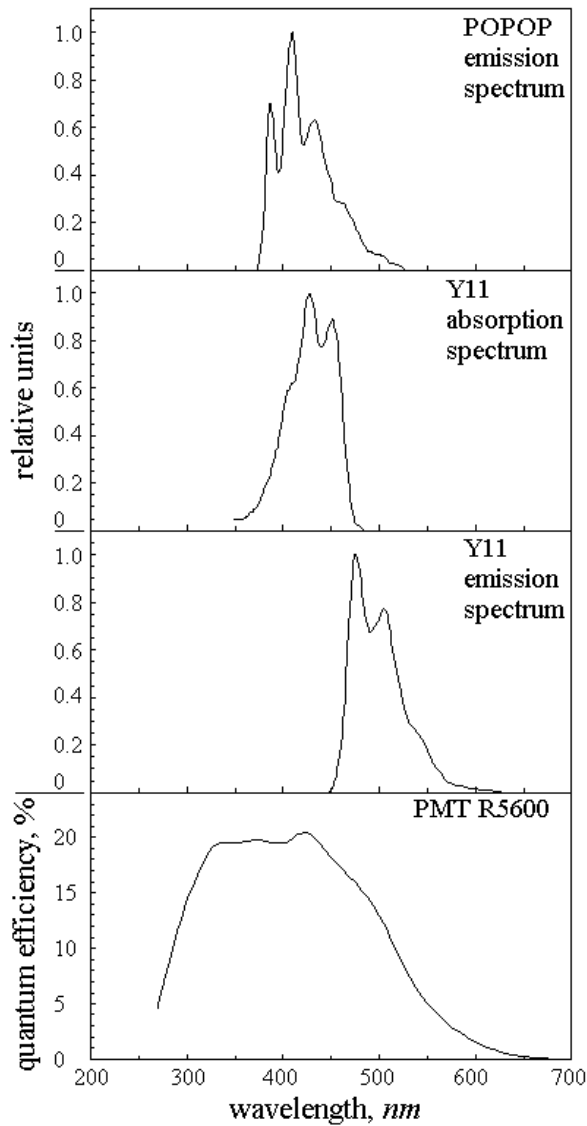

*Figure 8*. POPOP emission spectrum, fluorescence converter Y11 (K27 formulation) absorption and emission spectra and PMT R5600 quantum efficiency.

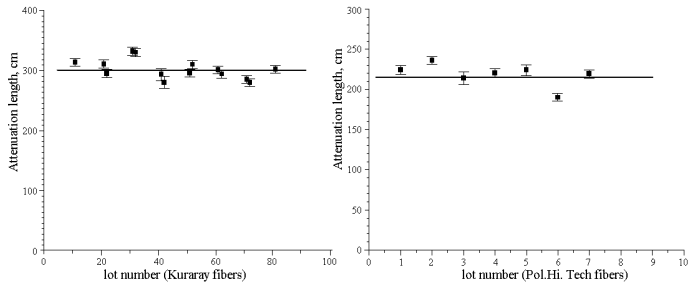

*Figure 9.* The average attenuation length of the Kuraray and Pol.Hi.Tech fibers as a function of lot number.

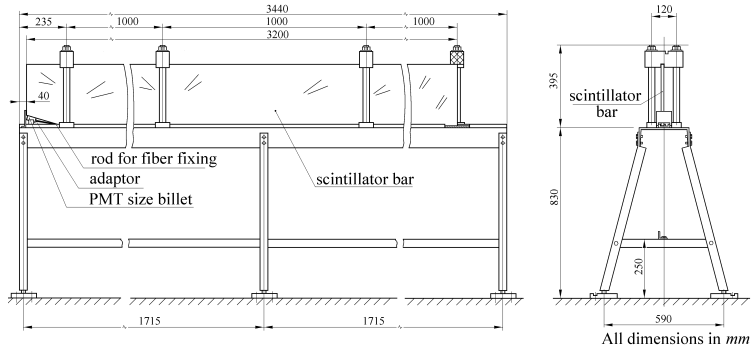

*Figure 10.* Assembly building-berth. Front and side view.

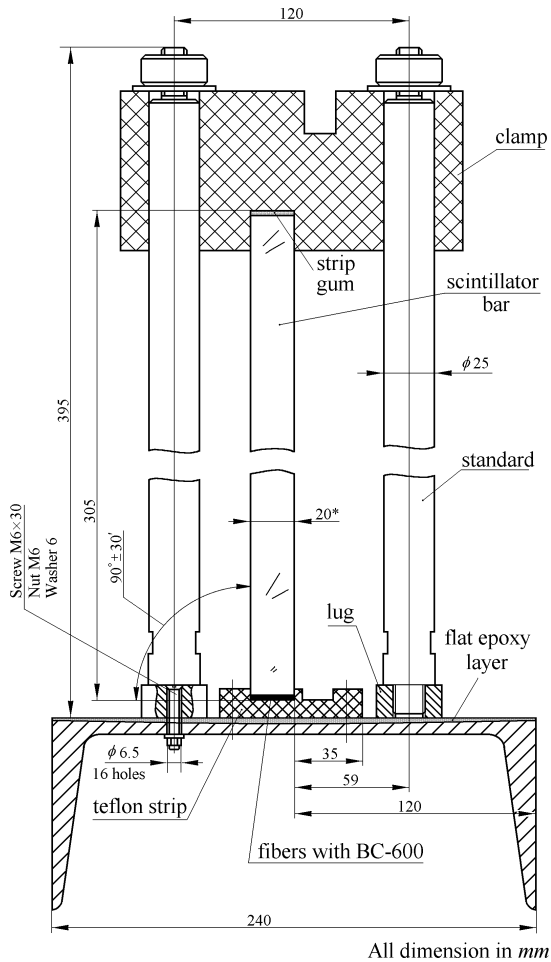

*Figure 11.* The clamping system.

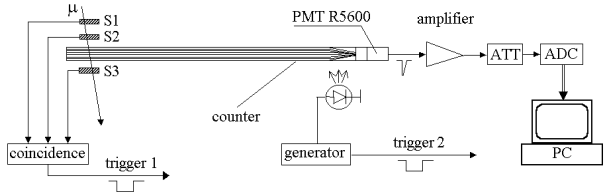

*Figure 12*. Block diagram of the setup used for the counters test.

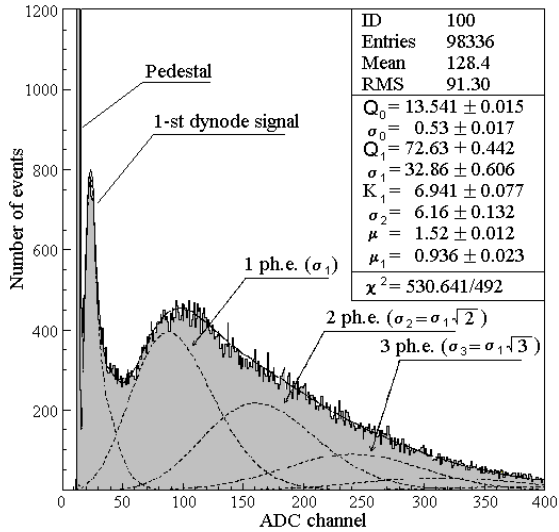

*Figure 13*. A typical deconvoluted LED spectrum of PMT R5600.

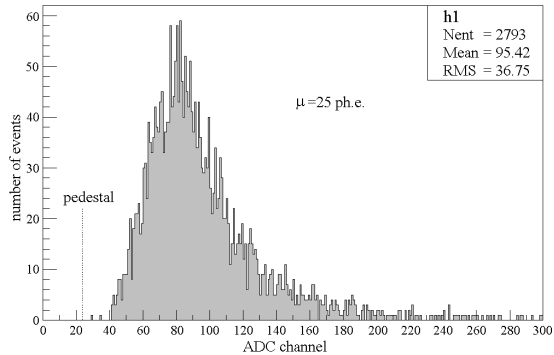

*Figure 14*. A typical cosmic muons spectrum taken from the far end of a counter. The dashed line shows the pedestal.

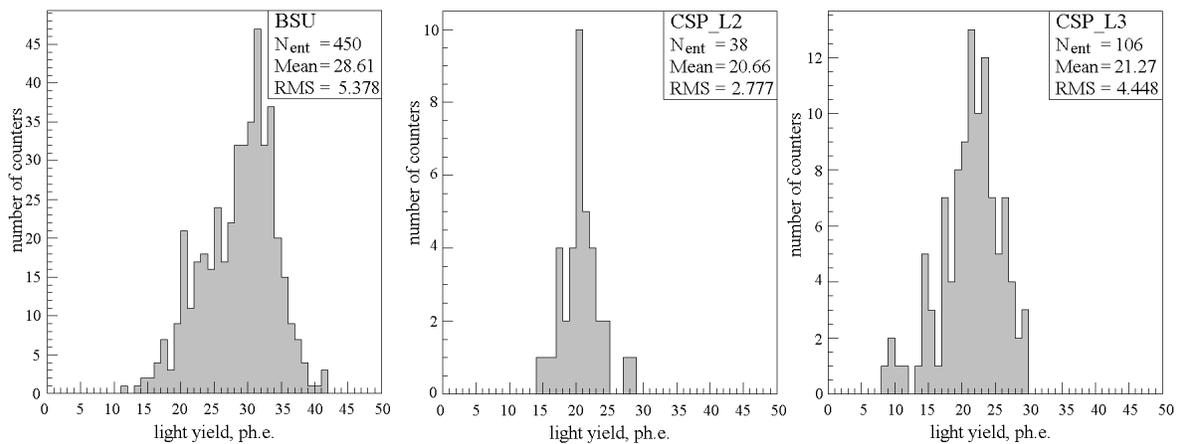

*Figure 15*. Distribution of counter light yield from the far end for BSU and CSP counters.